# RESPONSE TO CLAIMS MADE BY LUEBBERT AND PACHTER
## on 'MISCALIBRATION OF THE HONEYBEE ODOMETER' (2024)
### arXiv (https://arxiv.org/abs/2405.12998v1)

Mandyam V. Srinivasan[1], Jürgen Tautz[2] and Geoffrey W Stuart[3]

[1] Queensland Brain Institute, University of Queensland, Australia
[2] Faculty of Biology, University of Würzburg, Germany
[3] School of Psychological Sciences, University of Melbourne,
and School of Psychological Sciences, Macquarie University, Australia

This is a point-by-point response to a non-peer-reviewed document titled "The miscalibration of the honeybee odometer", authored by L. Luebbert and L. Pachter, that was published on arXiv on 8 May 2024 and subsequently publicised via social and mainstream media. The authors never contacted Srinivasan or Tautz personally with their queries, nor did they seek clarification. They do not work in this research area, and they have drawn several unjustified conclusions that reflect a limited understanding of the data they have examined.

Luebbert and Pachter have made several inaccurate interpretations of the published data, and even attempted to manipulate figures published by the laboratories of Srinivasan and Tautz in illogical and inaccurate ways to make unjustified allegations of "inconsistencies in results, duplicated figures, indications of data manipulation and incorrect calculation"

The authors have also falsely claimed that inadvertent data entry errors in two papers, which have since been corrected and published in *The Journal of Experimental Biology,* are examples of "data manipulation and duplication".

It is important to note that these minor corrections do not affect the conclusions of any of the studies, which remain firm and sound. These studies have also been replicated independently in many subsequent studies from other reputable laboratories.

### POINT-BY-POINT RESPONSE TO CLAIMS MADE BY LUEBBERT AND PACHTER

The Luebbert and Pachter document will hereafter be referred to as L&P.

### ALLEGATIONS OF MISCALIBRATION OF THE HONEYBEE ODOMETER

The questions raised by L&P on honeybee odometry reveal a lack of understanding of honeybee navigation. The following explanations demonstrate that the calibration provided of the honeybee odometer in Srinivasan et al (2000a) paper [8] is correct.

   (i)     Slopes and intercepts of waggle duration vs distance functions

In their Figure 1, L&P have erroneously calculated the slope of the waggle duration – versus distance plot by *including the intercept, which is incorrect,* as explained below.

In Srinivasan et al (2000a), the authors deliberately excluded consideration of the intercept in the waggle duration – versus distance plot (Fig. 2), because the intention was to calibrate the performance of the odometer *during cruising flight*. We calculated the *increase* in the waggle duration (ms) for a 1 m *increase* of the distance flown, to arrive at a slope of 1.88 ms/m. L&P have not understood our objective in calibrating the odometer. They have calculated the slope of the regression line incorrectly, as illustrated in Fig. 1.

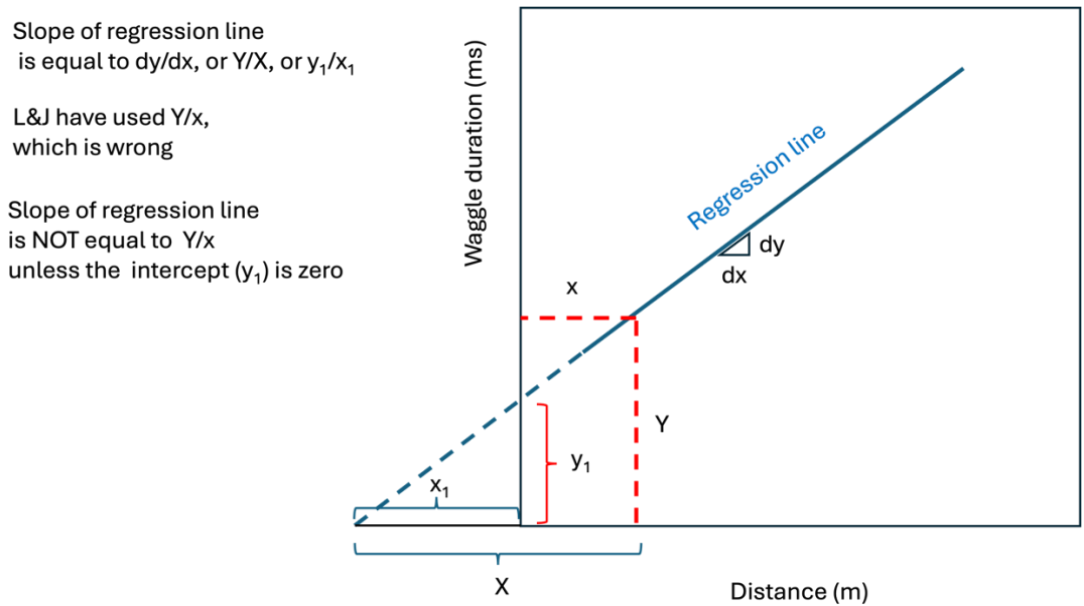

**Fig. 1** Illustration of erroneous calculation of slope by L&P

Visual inspection of Fig. 2 of the Schürch et al. (2019) paper [21] (ignoring the intercept), indicates a slope of approximately 1.4 m/ms. While our slope is higher (although likely not significantly so), one reason for this could be that our bees flew relatively short distances over an open meadow with sparse foliage and no trees. It is therefore likely that they flew closer to the ground, and thus experienced a greater magnitude of optic flow (image motion per unit distance travelled) compared to the bees in the Schürch et al. (2019) study. *Our findings are consistent with the 1995 and 1996 studies by Esch and Burns ([14], [15]), which demonstrated that bees signal a much lower distance in their dances when they fly at high altitudes.* Therefore, contrary to the claim made in the Schürch et al. (2019) study, there cannot be a 'universal' calibration of the honeybee odometer. The reading of the odometer (in terms of milliseconds per unit distance travelled) will depend heavily upon the environment through which the bee has flown: it will be lower if the flight is through an open meadow (for example) or over water (Tautz et al., 2004 [22]) and higher if it is through a densely wooded environment, or in a narrow tunnel (as in the Srinivasan et al. (2000a) experiments. The true calibration of the odometer must be based on *units of optic flow (the extent of visual image motion) experienced by the bee*. This is how we have arrived at the calibration figure 17.7 deg/ms. L&P are not aware of these nuances, because they do not work in this field.

On the relationship between waggle duration and image motion: L&P state "We noticed that the data does not support that 186 m of outdoor flight is encoded by a waggle duration of 350 ms unless the y-intercept (96 ms) is ignored." But, as we have pointed out above, the y-intercept *must* be ignored if one wishes to obtain a true in-flight calibration of the odometer. Their argument is therefore incorrect.

The arguments that L&P make in their Figures 1 and 2 to support their claims are invalid, for the reasons explained above.

(ii) Quantization of waggle duration by frame rate (40ms)

The analysis provided by L&P in their Table S1 is flawed. In this table, which provides an analysis of the honeybee waggle dance data, the third set of numbers in the rightmost column [(Mean*n)/40 for each n] displays non-integer values simply because the values of the mean waggle durations, shown in the leftmost column in milliseconds, have been rounded to one decimal place. Table 1 (below) shows a calculation of (1) the nearest mean waggle durations representing multiples of 40ms, and (2) their rounded values (rounded to one decimal place).

| Condition | Hive | Duration | Num | Exact* | Rounded | Match |
|---|---|---|---|---|---|---|
| 60m | H1 | 217.5 | 92 | 217.3913 | 217.4 | No (+0.1) |
| 110m | H1 | 283.0 | 93 | 283.0107 | 283.0 | Yes |
| 110m | H2 | 312.4 | 181 | 312.4862 | 312.5 | No (-0.1) |
| 150m | H1 | 390.9 | 92 | 390.8696 | 390.9 | Yes |
| 190m | H1 | 441.2 | 65 | 441.2308 | 441.2 | Yes |
| 225m | H2 | 514.4 | 345 | 514.4348 | 514.4 | Yes |
| 340m | H2 | 733.3 | 222 | 733.3333 | 733.3 | Yes |
| 350m | H1 | 762.6 | 87 | 762.7586 | 762.8 | No(-0.2) |
| Tunnel Exp 2 | H2 | 528.8 | 216 | 528.8889 | 528.9 | No(-0.1) |
| Tunnel Exp 4 | H2 | 441.2 | 138 | 441.1594 | 441.2 | Yes |

**Table 1** Exact* represents the mean required to produce an exact multiple of 40ms for mean duration x number of waggles (Num). Note that for mismatches there was no exact value within the range Duration +/- .05.

The rounded values match the values given in Table 1 of Srinivasan et al (2000a) [8] in 6 out of 10 cases. There is a mismatch of magnitude 0.1 ms in 3 cases, and of magnitude 0.2 ms in one case. These mismatches are likely due to minor inconsistencies in the rounding operations (rounding up instead of down or vice-versa), or dropping one or more trials post video analysis. The four mismatches (with a maximum magnitude of 0.2 ms) are very small in relation to the value of the measured quantity (waggle duration), which ranges from 217 ms to 762 ms. They have a negligible effect on the computed linear regression equation and correlation coefficient, as shown below for the Exact* values and the Rounded values of the waggle durations given in Table 1 above. The procedures for computing the linear regression equation and the correlation coefficient (R) are described in Appendix B.

For Exact* values: R=0.998 (*Excellent agreement with published value of 0.998*)
Regression equation: Tau=95.89 + 1.88*d
(*Close agreement with the published regression equation Tau=95.91 + 1.88*d*)

For Rounded values: R=0.998 (*Excellent agreement with published value of 0.998*)
Regression equation: Tau=95.90 + 1.88*d
(*Close agreement with the published regression equation Tau=95.91 + 1.88*d*)

L&P's computation of the first and second set of numbers in the rightmost column of their Table S1 is not logical: there is no expectation for these numbers to be integers. The video camera was *not* used to count the number of bees, the number of dances, or the number of waggle phases. These numbers were obtained from manual observation of the recorded video dances; they have nothing to do with the frame rate of the video, which is irrelevant in this context.

The basic findings from examining the visual cues that underlie odometry in honeybees – as reported in [1] and [2] - have been replicated in other studies, not only in our laboratory (e.g. [7],[8],[17],[18],[22]), but in other laboratories as well (e.g. [10],[11],[19],[20]).

The journal *Science* has now published a commentary about L&P's allegations in their *News* Section on 5 July 2024. We provided a rebuttal to this commentary, which they published below their commentary as a 'Response' on 3 August. On 10 August 2024 *Science* revised their commentary to clarify that "Valda Vinson, executive editor of the *Science* family of journals, says, '*Science* editorial has evaluated the concerns and stands by the (Srinivasan et al., 2000) published paper, at this time.'"

## ALLEGATIONS OF DATA DUPLICATION AND POTENTIAL MANIPULATION
### (Figures 3, 4, Supplementary Figs. 1 & 2, and Table 1 of L&P)

In their Figs. 3, 4, and Supplementary Figure 1, the discussion by L&P of our data and their claims of data manipulation are invalid. In Fig. 3 and Supplementary Figure 1, their claims of identical data being reported for different experimental conditions have already been addressed and the corrections published in the *Journal of Experimental Biology.* Further details are provided in Appendix A.

In L&P's Supplementary Figure 2, they have manipulated our figures in illogical and inaccurate ways to make the initial data points in different curves look alike. In Fig. 9 of the 2000 *Biological Cybernetics* paper (Srinivasan et al., 2000b) [16], the scales for the horizontal axes (Frame number) and the vertical axes (Horizontal distance travelled) are not identical in the four panels (a, b, c, and d). The maximum values along the 'Frame number' axis are: 25 for (a), 15 for (b), 30 for (c), and 30 for (d). The maximum values along the 'Horizontal distance travelled' axis are: 60 for (a), 50 for (b), 60 for (c) and 50 for (d). L&P have compressed or stretched (not "skewed") the horizontal and vertical scales inaccurately to try to match the data points in the initial parts of the curves.

Again, in L&P's Supplementary Fig. 2, if one looks at the top left-hand panel - where they compare the curves for land 16, land 18 and land 22 – it is evident that the vertical scale markings do not match up. The red 60 is higher that the grey 60; the red 50 is higher than the grey 50; and so on. The horizontal scales (Frame number) are also not exactly matched: In the same panel, the grey 25 is not exactly half-way between the green 20 and the green 30. Therefore, in reality, the six labelled points do not match. Furthermore, while these six labelled points overlay each other, the other points in the curves do not match at all, even after this process of stretching and compression. In addition, the three curves have different numbers of data points (land 16: 25 points; land 18: 26 points; land 22: 27 points). There is therefore no basis for the suggestion of manipulation of these data sets. The close similarity of some data points (or disparity in others) is irrelevant. These are all distinct data sets that have not been duplicated or manipulated.

The above considerations also apply to the stretched and compressed graphs displayed in the upper right and lower left panels of L&P's Supplementary Figure 2. Again, L&P have compressed or stretched the graphs inaccurately to try to match the initial five data points (land 09 and land12 in the lower panel) and the final five data points (land 09, land 12 and land 21in the upper right panel).

Close inspection of the lower left panel reveals that only four of the five labelled data points match closely. Close inspection of the lower left panel again reveals that only one of the five labelled data points matches closely. Furthermore, the three curves carry different numbers of data points: (land 09: 20 points; land 12: 21 points; land 21: 19 points). Again, there is no basis for the suggestion of manipulation of these data sets.

The suggestions put forth by L&P for replication or manipulation of data in this way are not logical.

Fig. 4A (rightmost panel) is another example of L&P's claim that "the data in Figures 8A and 9B are highly similar, as evident when Figure 9B is slightly skewed and flipped vertically". This claim is extraordinary. For one thing, 8A and 9B show different relationships: 8A plots Height versus Frame number, while 9B plots Horizontal distance travelled versus frame number. The vertical axes represent completely different variables, and there is no rational reason for flipping one set of data points to match another set. Furthermore: (i) Figure 8A contains 20 data points, whereas Figure 9B contains 12 data points; (ii) a close inspection of the flipped and overlapped curves reveals that, contrary L&P's claim, points 2, 4, 7 and 10 do not overlap. The data sets for these two curves are clearly different: they have not been replicated. They are from two different landing trajectories (09 and 17). Attempting to claim data duplication or manipulation by flipping the data points tracking a variable in one graph *upside down* to try to match (inaccurately) the data points in another graph that is tracking a *different* variable, and involves a *different* landing trajectory, is an exercise that is illogical and fundamentally flawed.

Figure 3B in [16] does indeed replicate the data from Figure 5A in [1]. We have stated in [16] that "A preliminary study of landing bees was published in [1]". But Srinivasan notices now that the X axis of Figure 5A has been incorrectly labelled. The correct X- axis labelling is provided in Figure 3B in [16]. The reason for this discrepancy is most likely that the horizontal distance travelled was erroneously multiplied by the video frame (25 frames/s) in labelling the X-axis labelling of Figure 5A in (1). This correction has been submitted to the *Journal of Experimental Biology* for publication as an erratum. This error does not affect any of the analyses, results, or conclusions in [1] or [16]. There was no deliberate duplication or manipulation of the data in any of the graphs presented in [1] or [16].

*Regarding L&P's claims in their Table 1:*

> The claim of data duplication and potential manipulation across six of our papers dating from 1996 to 2004 is invalid.

- The queries in relation to the 1996 and 1997 *JEB* papers have already been answered above.

- We see no problem with the re-use of data from certain control experiments conducted in the 1996 *JEB* study in the 1997 *JEB* study: At the end of the Introduction section of the 1997 paper we have stated explicitly that "A preliminary account of this work is given in Srinivasan et al. (1996)."

- The data from Fig. 3b,c in the 2000 *Biological Cybernetics* paper is indeed a replicate of the data in Fig. 5A of the 1996 *JEB* paper. We have acknowledged this in the *Biological Cybernetics* paper by stating that "A preliminary study of landing bees was published by Srinivasan et al. (1996)".

- There is no data inconsistency in the methods or results presented in Srinivasan et al. (*Science,* 2000) with those of other studies, as discussed above under the section on honeybee odometry. This is an erroneous conclusion by L&P arising from their lack of understanding of bee navigation.

- It is claimed that, in the 2004 paper on *Landing Strategies in Honeybees and Applications to UAVs*, "Figures 11 and 12 are identical." There is no Figure 11 or 12 in this paper, which contains only 4 figures.

- It is claimed that the 2004 paper on *Landing Strategies in Honeybees and Applications to UAVs* "Republishes problematic data from the 2000 *Biological Cybernetics* paper." We have already shown above that there are no problems with the data in the 2000 paper. Secondly, the 2004 paper is a review of work done in the laboratory; it is not original research. It therefore covers data published in earlier papers. This is acknowledged and clearly stated in each figure of this review paper

- It is claimed that the 2004 paper on *Visual Motor Computations in Insects* "Republishes problematic data from the 2000 *Biological Cybernetics* paper and the 1997 paper". The 2000 and 1997 papers have already been discussed in detail above. We have already shown above that there are no problems with the data in the 2000 paper. Again, this 2004 paper is a review of work done in the laboratory; it is not original research. It therefore covers data published in earlier papers. This is acknowledged and clearly stated in each figure of this review paper.

## CORRELATION OF .99 BETWEEN R2 =0.99 REGRESSIONS AND HONEYBEE NAVIGATION (Table 2 in L&P)

On the question of a high linear regression coefficient reported for the graph in Fig. 2 of Srinivasan et al (2000a) [8]: This coefficient was computed using just the mean value of the data for each outdoor flight distance, given in Table 1 of the paper. Recalculation of the regression coefficient yields a value of 0.998, which is exactly the published value of 0.998 - **there is no error**. Details are provided in Appendix B. *It is also worth noting that similarly high values of this regression coefficient have also been observed in the classic studies of honeybee waggle dance behavior by von Frisch (1946) and Wenner (1962), as reported in Schürch et al. (2013) [25].*

### CALCULATIONS OF REGRESSION COEFFICIENTS AND REGRESSION LINE EQUATIONS

The raw data collected in the studies from the cited papers (published in 2000, 2004, 2005, and 2010) are unfortunately no longer available. However, we can respond to the specific question relating to the comment on the high linear regression coefficients (R). In the graphs that show individual markers, the markers represent individual data points, and the regression coefficients were computed using them as single measurements. In the graphs that include standard deviations, the regression coefficients were computed by using only the **mean** value of the ordinate for each (independent) value of the abscissa. This is the reason for the rather high values of the computed regression coefficients.

We have checked the reported values of the linear regression coefficients and the equations of the regression lines by manually digitizing the data points in the published figures, and by using two different methods to compute the coefficients and the linear regression equations. The results (shown in detail in Appendix B below) are in very good agreement with the published values. To the best of our knowledge, there are no errors in the calculation of the regression coefficients. However, our calculation checks indicate that the $R^2$ values reported in some of the papers in question are actually R values, not $R^2$ values (correlation coefficient, not coefficient of determination), as detailed in Appendix B. This is a typographical error, which will be corrected and published as errata in the relevant journals.

There are, of course, inevitable minor discrepancies arising from errors in the manual digitization process. In the case of Fig. 2 in Srinivasan et al. (2000a) and Fig. 1b in Zhang et.al. (2005) [23] manual digitization was not necessary as the data are supplied in Table 1 (Srinivasan et al. (2000a)), and in the inset to Fig. 2b (Zhang et.al. (2005)), respectively.

*In conclusion, many of the questions posed by L&P stem from (a) a lack of familiarity with this research field and knowledge of how relevant data is collected and analysed; and (b) prior assumptions of dishonesty and data manipulation. These allegations are unjustified and emphatically refuted in the explanations given above.*

L&P have also claimed, via simulation, that the reported $R^2$ value in Srinivasan et al. (2000) is not reproducable from the reported means and standard deviations in Table 1 of that paper. Their simulations suffer from an irrecoverable limitation and two serious flaws. A full account of these problems is given in Stuart (2024) (https://arxiv.org/abs/2408.07713).

Briefly, the limitation is that the waggle durations in Srinivasan et al. (2000) do not represent independent observations. The data follow a hierarchical structure with four levels (hive, bee, dance and waggle). To reproduce the original $R^2$, they should be simulated as such (including covariances within and between levels), but this would require the raw data, which is no longer available.

The first and most serious flaw is that unlike Srinivasan et al. (2000), who averaged over all waggles at a given distance prior to regression, L&P (in their first simulation) used bees as "replicates". However, instead of averaging over dances and waggles for each bee, they simulated a single waggle duration, taken from the overall waggle distribution at that distance. In effect, they confused the *standard error* of the mean (for each bee), for the *standard deviation* of the full set of waggle durations. This elementary statistical error markedly increased the variance of the supposed mean waggle durations per bee prior to regression, making it nearly impossible to reproduce the original $R^2$ value.

The second flaw is that L&P used the observed means in Table 1 of Srinivasan et al. (2000) as the starting point of their simulations. This approach confuses sample means with underlying population means, which all available evidence suggests follow a linear function. By using this approach, L&P "doubled up" on sampling error, again increasing the variance of mean waggle duration, thereby attenuating their simulated $R^2$ values.

As shown in Stuart (2024), when these flaws are addressed, the reported $R^2$ of Srinivasan et al. (2000) is within the simulated range and is consistent with $R^2$ values in other studies who averaged over many waggle durations per bee prior to regression.

# APPENDIX A

The results in all the figures in Srinivasan et al. (1996) [1] and Srinivasan et al. (1997) [2] were obtained from work in Srinivasan's laboratory at the Australian National University in Canberra - except for Fig. 1 in Srinivasan et al. (1996), which was from work in Prof. Thomas Collett's laboratory at the University of Sussex.

The results shown in Figs. 2A (stripes), 2C and 3(B,C) of Srinivasan et al. (1997) should be identical to those reported in Figs. 7, 8A and Figs. 8(C,D), respectively, in Srinivasan et al. (1996). These experiments were not repeated for the 1997 study, and this has been acknowledged the earlier work in the last sentence of the Introduction section in Srinivasan et al. (1997): "A preliminary account of this work is given in Srinivasan et al. (1996)". In those days, authors were not required to obtain copyright permission for reproducing figures.

Srinivasan has looked through all of his storage media. Unfortunately, it has not been possible to find raw data, analyses, or figure plots dating back to 1996 and 1997 to determine the reasons for these discrepancies. This information cannot be obtained from the other collaborators in these studies either, because their contact details are out of date.

However, Srinivasan has examined the information provided in subsequent publications ([3], [4], [5] and [6]), which have reviewed this (and other) work from the laboratory. These subsequent publications indicate consistently that the length of the tunnel was 3.20m, as shown in Fig.1 of Srinivasan et al (1997), and not 3.35m, as shown in Fig. 6 of Srinivasan et al (1996). So, the 3.35m length shown in Fig. 6 of Srinivasan et al (1996) was definitely a typographical error.

With regard to the discrepancy between the stated width of the narrow tunnel [7cm in Fig. 3B of Srinivasan et al (1997) and 11cm Fig. 8C in Srinivasan et al (1996)]: The original information on this is no longer available but it is very likely that the correct width was 11cm, because the aim of that experiment was to compare the searching distance in a wide tunnel (22cm), with that in a tunnel half as wide (11cm), to see if the results conformed with the expectation that the searching distance should be twice as long in the wider tunnel. This indeed turned out to be the case, when comparing the locations of the peaks of the searching distributions in the two tunnels. Besides, as honeybees are very reluctant to fly continuously through tunnels as narrow as 7 cm, it is very likely that the width of the narrow tunnel was not 7 cm (typographical error), but 11cm.

On the question of the discrepancy in the sample sizes reported for one particular experiment using the 22 cm wide tunnel [n=56 in Fig. 3B (1997), and n=88 Fig. 8C (1996)]: Srinivasan does not have the original data and is unable to check these numbers in the subsequent review articles, because they do not include the sample sizes. But the best guess would be that the number n=88 given in the earlier publication (1996) is the correct figure, and the number n=56 provided for the same experiment in the subsequent publication (1997) is an error of numerical entry.

On the question of the search distribution in shown in Fig. 7 (1997) for the feeder at unit 9 being identical to the distributions shown in Figs. 2, 3 and 5, but showing a sample size of N=71 rather than N=121: Again, Srinivasan does not have the original data, but it appears that Fig. 7 does not contain the correct graph for the searching distribution with the landmark positioned at Unit 9, which was a separate experiment from those described in Figs 2, 3 and 5.

The 'still air' control in Fig. 3A and the control Fig. 3B show the results of the same control experiment but there is a discrepancy in the sample sizes. It is impossible, at this stage, to specify which sample size is the correct one (N=19 or N=35).

On behalf of all the authors of the 1996 and 1997 papers, Srinivasan sincerely apologizes for these errors and takes sole responsibility for them. However, these corrections do not alter the conclusions of the two papers in any way. The basic findings from examining the visual cues that underlie odometry in honeybees – as reported in [1] and [2] - have been replicated in other studies, not only in Srinivasan's laboratory (e.g. [7],[8],[9]), but in other laboratories as well (e.g. [10],[11]).

The corrections to [1] and [2] were summarized in two notices published by the *Journal of Experimental Biology* on 15 June 2024 ([12],[13]).

L&P state that "The spatial distributions of the honeybee search shown in Figure 6C in the 1996 paper and in Figure 1C in the 1997 paper differ, even though the rest of the data is identical." L&P have not understood what these figures represent. They are cartoon diagrams illustrating the procedure for measuring the spatial distribution of the honeybee search. They do not represent data. Whether these two cartoon diagrams are identical, similar or very different is irrelevant.

# APPENDIX B

For the papers in question, Srinivasan has checked the published values of the coefficients of linear regression, R, and the linear regression equations by repeating the calculations after manually digitizing the points in the graphs of a few sample figures.

As noted in the main text, the $R^2$ values reported in some of the papers in question are actually R values (correlation coefficient, not coefficient of determination).

For each case, R was computed using two methods:

(a) From the Pearson's correlation coefficient, using the coordinates of the $n$ points on the graph $(x_i, y_i)$ $(i = 1, 2, ... n)$:

$$R = \frac{I_{XY}}{I_{XX} I_{YY}} \quad (1)$$

where $I_{XY} = \sum_{i=1}^{n}(x_i - \bar{x})(y_i - \bar{y})$, $I_{XX} = \sqrt{\sum_{i=1}^{n}(x_i - \bar{x})^2}$, $I_{YY} = \sqrt{\sum_{i=1}^{n}(y_i - \bar{y})^2}$, $\bar{x} = \left(\frac{1}{n}\right)\sum_{i=1}^{n} x_i$, and $\bar{y} = \left(\frac{1}{n}\right)\sum_{i=1}^{n} y_i$.

(b) From Matlab's linear regression fitting model routine 'fitlm'.

The linear regression equation was computed using

(c) Matlab's polynomial fitting function 'polyfit' for a first-order fit;

and

(d) Matlab's linear regression fitting model routine 'fitlm'.

**Check of linear regression coefficients published in various papers**

**Evangelista et al. 2010** [24]
Evangelista, C., Kraft, P., Dacke, M., Reinhard, J. and Srinivasan, M.V., 2010. The moment before touchdown: landing manoeuvres of the honeybee Apis mellifera. *Journal of Experimental Biology*, *213*(2), pp.262-270.

*Fig. 5*
*Body-platform angle versus platform tilt (filled circles, 17 manually digitized points):*
*Published results:*
$R^2$= 0.99; → R= 0.995. Regression equation: Body-platform angle = 0.82*tilt -11.49

*Check of calculations:*
From equn (1): R=0.996. *(Good agreement)*

From Matlab polynomial fit model (for first-order fit):
Regression equation: Body-platform angle = 0.82*tilt -11.93 *(Good agreement)*

From Matlab linear regression model (ML):

$R^2$ = 0.985; → R= 0.992. *(Good agreement)*
Regression equation: Body-platform angle= 0.82*tilt -11.93 *(Good agreement)*

*Body-horizontal angle versus platform tilt (open circles, 18 manually digitized points):*
*Published results:*
$R^2$ = 0.92; → R= 0.96. Regression equation: Body-horizontal angle = 0.16*tilt -13.74

*Check of calculations:*

From equn (1): R=0.96. *(Excellent agreement)*

From Matlab polynomial fit model (for first-order fit):
Regression equation: Body-horizontal angle= 0.16*tilt -13.34 *(Good agreement)*

From Matlab linear regression model (LM):
$R^2$ = 0.92; → R= 0.96. *(Excellent agreement)*
Regression equation: Body-horizontal angle= 0.16*tilt -13.34 *(Good agreement)*

**Srinivasan et al. 2000a** [8]
Srinivasan, M.V., Zhang, S., Altwein, M. and Tautz, J., 2000a. Honeybee navigation: nature and calibration of the" odometer". *Science*, 287(5454), pp.851-853.
Fig. 2 (8 points, tabulated in Table 1):
*Published results:*
R=0.998. Regression equation: Tau=95.91 + 1.88*d

*Check of calculations:*

From equation (1): R= 0.998. *(Excellent agreement)*

From Matlab polynomial fit model (for first-order fit):
Regression equation: Tau=95.96 + 1.88*d *(Good agreement)*

From Matlab linear regression model (LM):
$R^2$ = 0.997; → R= 0.998. *(Excellent agreement)*
Regression equation: Tau= 95.96+1.88*d *(Good agreement)*

**Srinivasan et al. 2000b** [16]
Srinivasan, M.V., Zhang, S.W., Chahl, J.S., Barth, E. and Venkatesh, S., 2000b. How honeybees make grazing landings on flat surfaces. *Biological Cybernetics*, 83, pp.171-183.
Fig. 6a (land01): 14 manually digitized points
*Published results:*
R=0.14. Regression equation: Vd=0.82*h +14.80

*Check of calculations:*

From equation (1): R= 0.14. *(Excellent agreement)*

From Matlab polynomial fit model (for first-order fit):
Regression equation: Vd=0.81*h + 14.81 *(Good agreement)*

From Matlab linear regression model (LM):

$R^2$ = 0.0198; → R= 0.14. *(Excellent agreement)*
Regression equation: Vd=0.81*h + 14.81 *(Good agreement)*

Fig. 6d (land17): 11 manually digitised points

*Published results:*
R=0.94. Regression equation: Vd=5.61*h -9.42

*Check of calculations:*

From equation (1): R= 0.94. *(Excellent agreement)*
From Matlab polynomial fit model (for first-order fit):
Regression equation: Vd=5.59*h -9.12 *(Good agreement)*

From Matlab linear regression model (LM):
$R^2$ = 0.887; → R= 0.94. *(Excellent agreement)*
Regression equation: Vd=5.59*h - 9.12 *(Good agreement)*

**Zhang et al. 2005** [23]
Zhang, S., Bock, F., Si, A., Tautz, J. and Srinivasan, M.V., 2005. Visual working memory in decision making by honey bees. *Proceedings of the National Academy of Sciences*, 102(14), pp.5250-5255

The $R^2$ values reported in Fig. 1 b and Fig. 1c are actually R values (correlation coefficient, not coefficient of determination).

Fig. 1b (8 points, tabulated in figure inset)

*Published results:*
R= 0.985; Regression equation: t=0.018*$X_d$ + 0.421

*Check of calculations:*

From equation (1): R= 0.99. *(Good agreement)*

From Matlab polynomial fit model (for first-order fit):
Regression equation: t=0.018*$X_d$ + 0.420 *(Good agreement)*

From Matlab linear regression model (LM):
$R^2$ = 0.985; → R= 0.99. *(Good agreement)*
Regression equation: t=0.018*$X_d$ + 0.420 *(Good agreement)*

Fig. 1c: 8 manually digitised points

*Published results:*
R= 0.999; Exponential curve fit: y=0.4 exp(-0.53t) + 0.50

The regression for a fitted exponential curve was obtained using the procedure described in [26, 27]. The exponential function was fitted using y=b*exp(-ct) + 0.50 and determining the values of the parameters b and c that produced the highest value of $R^2$. Srinivasan is not sure if this was the procedure used in the study, because he was not involved in the data analysis part of this paper. The procedure described above yields the function y=0.4 exp(-0.53t) + 0.50 (red curve, Fig. B1). This function exactly matches the solid curve shown in Fig. 1c. It appears that there is a typographical error in the function specifying the curve in Fig. 1c, which is reproduced as the blue curve in Fig. B1. The parameter (b) multiplying the exponent should be 0.40, not 0.48.

The result of the regression calculation yields $R^2$ = 0.911; → r = 0.954. The reason for the discrepancy between the published value of 0.999 and the newly computed value of 0.954 is either (a) an error of numerical insertion, or (b) the original calculation of the regression was performed using a different method, information about which is no longer available. In any case, the observation made in the paper is simply that the curve decays approximately exponentially. Even if this decay is not strictly exponential, this does not change the conclusions of the paper in any way.

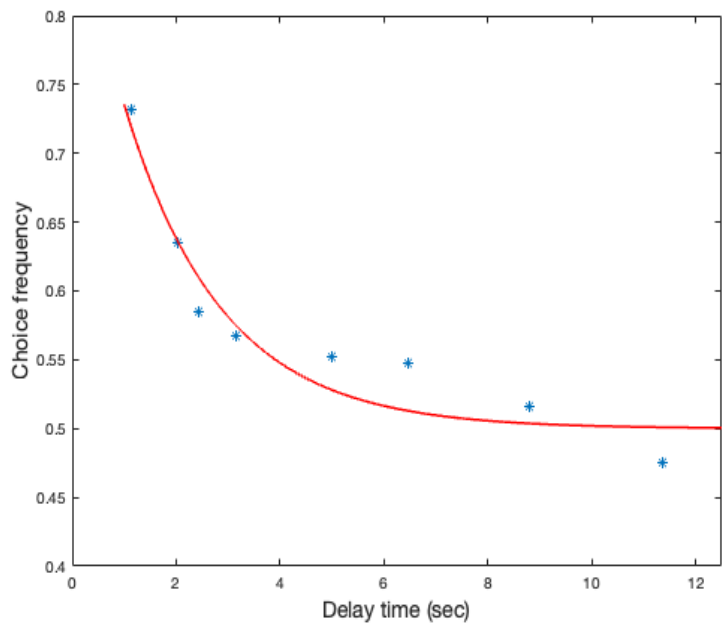

*Fig. B1 Exponential fitting function y=b\*exp(-ct) + 0.5 recalculated for b=0.40, c= 0.53 (red curve). This exactly matches the solid curve shown in Fig. 1c of Zhang et al. (2005) [23].*

**Tautz et al. 2004** [22]
Tautz, J., Zhang, S., Spaethe, J., Brockmann, A., Si, A. and Srinivasan, M., 2004. Honeybee odometry: performance in varying natural terrain. *PLoS Biology*, *2*(7), p.e211.
(Note: Published values are for $R^2$, not R)

Fig. 2a:  16 manually digitized points

*Published results:*
$R^2$=0.9613. Regression equation: Y=1.303\*X +202.4

*Check of calculations:*
From equation (1): R= 0.9804  → $R^2$= 0.9612. (*Good agreement*)

From Matlab polynomial fit model (for first-order fit):
Regression equation: Y=1.296*X +219.0 (*Good agreement*)

From Matlab linear regression model (LM):
$R^2$= 0.961. (*Good agreement*)
Regression equation: Y=1.296*X +219.0 (*Good agreement*)

Fig. 3a: 16 manually digitized points

*Published results:*
$R^2$=0.9770. Regression equation: Y=1.431*X +168.6

*Check of calculations:*

From equation (1): R= 0.9888 → $R^2$= 0.978.  (*Good agreement*)

From Matlab polynomial fit model (for first-order fit):
Regression equation: Y=1.446*X +174.1 (*Good agreement*)

From Matlab linear regression model (LM):
$R^2$= 0.978 (*Good agreement*)
Regression equation: Y=1.446*X +174.1 (*Good agreement*)

*****